\documentclass[reqno, 11pt]{amsart}
\usepackage{amsfonts }
\usepackage{graphicx}
\usepackage{epsfig}
\usepackage{epstopdf}

\theoremstyle{plain}

\begin{document}
\markboth{J. K. Singh $\cdot$ Sarita Rani}{Self Interacting Brans Dicke Theory}
\noindent{\bf {The Bianchi type-V Dark Energy Cosmology in Self Interacting Brans Dicke Theory of Gravity}}
\vskip0.4in
\begin{footnotesize}

\noindent{\bf J. K. Singh\textsuperscript{$*$} $\cdot$ Sarita Rani\textsuperscript{$\dagger$}}\\
\noindent Department of Mathematics, Netaji Subhas Institute of Technology\\
University of Delhi\\
Sector-3, Dwarka, New Delhi-110078, India\\
\noindent \textsuperscript{$*$}jainendrrakumar@rediffmail.com\\
\noindent \textsuperscript{$\dagger$} sarita\_maths@yahoo.co.in\\

\vskip0.4in
\noindent{\bf Abstract} This paper deals with a spatially homogeneous and totally anisotropic Bianchi type-V cosmological model within the framework of self interacting Brans Dicke theory of gravity in the background of anisotropic dark energy (DE) with variable equation of state (EoS) parameter and constant deceleration parameter. Constant deceleration parameter leads to two models of universe, i.e. power law model and exponential model. EoS parameter $\omega$ and its existing range for the models is in good agreement with the most recent observational data. We notice that $\omega$ given by (\ref{30c}) i.e $\omega(t)=log(k_1t)$ is more suitable in explaining the evolution of the universe. The physical behaviors of the solutions have also  been discussed using some physical quantities. Finally, we observe that despite having several prominent features, both of the DE models discussed fail in details.

\end{footnotesize}
\textbf{PACs numbers}: 4.50.Kd, 98.80.-k
\vskip0.3in
\noindent{\bf 1 Introduction.}
\vskip0.2in
\noindent Current accelerated expansion in the universe verified by numerous ways such as high red-shift supernovae Ia (SNeIa) experiments \cite{per3,rie1}, cosmic microwave background (CMB) radiations \cite{spe1,spe2}, large scale structure LSS \cite{teg},  WMAP \cite{ben}, SDSS \cite{teg}, galactic cluster X-rays emission \cite{all}, weak lensing \cite{jai} etc. has been proven a milestone in exploring the mysteries of universe. After this revolution, major hurdle still left over is to describe the theoretical aspects behind it and to search the component responsible for it. As per the latest PlANK $2013$ results \cite{pla}, our universe contains $4.9\%$ ordinary i.e baryonic matter, $26.8\%$ dark matter and $68.3\%$ dark energy. Noticing this amazing and wonderful composition of the universe, dark energy, a cosmic fluid with negative pressure has been proved to be the most trusted and accepted candidate accounting for the accelerated expansion in the universe. Still the knowledge of exact nature of dark energy is in dark, therefore, we further need to explore it which observationally seems to be possible by determining its equation of state (EoS) parameter $p=\omega(t)\rho$, where $p$ is the fluid pressure and $\rho$ is the energy density \cite{car3} but yet exact value of EoS parameter is not accurately determined. Normally the value of the EoS parameter is taken a constant \cite{kuj,bar} as $-1$, $0$, $\frac{1}{3}$ and $1$ for a vacuum fluid, dust fluid, radiation and stiff-matter dominated universe, respectively but more generally, its not always constant and may be considered as a function of cosmic time or redshift \cite{jim,das,rat}. Recently, Sahni and Starobinsky \cite{sah2} developed techniques for regeneration of $\omega(t)$ from experimental data and further Sahni et al. \cite{sah3} analysed the data obtained to determine it as a function of cosmic time. The simplest case of DE is the vacuum energy with $\omega=-1$ popularly known as the cosmological constant. Some of the others time dependent EoS parameter models are quintessence models involving scalar fields with  $-1<\omega\leq-\frac{1}{3}$ \cite{car3}, phantom phase with $\omega<-1$ and quintom model inheriting characteristics of both i.e. quintessence as well as phantom, crossing the phantom divide line $\omega=-1$ and have time dependent EoS parameter. Cosmological models with variable EoS parameter have also been studied in Kaluza-Klein metrics and wormholes \cite{rah}. Recently, numerous time dependent forms of $\omega$ have been used for variable cosmological constant models \cite{muk}. Bamba et al \cite{bam} reviewed various dark energy cosmologies and showed that that almost all dark energy models are equivalent and finally approach to $\Lambda$CDM i.e. cosmological constant model, consistent with the observational data. They explained observational tests constraining the current cosmic acceleration and also explored general equation of state EoS. Several other authors, Akarsu and Kilinc \cite{aka}, Singh and Sharma \cite{sin1,sin2,sin3} have discussed dark energy models with a variable EoS parameter. Singh and Rani \cite{sir2} have studied  Bianchi type-III  cosmological models with modified Chaplygin gas (MCG) having variable equation of state $p=A_1\rho-\frac{A_2}{\rho^\alpha}$ where $0\leq A_1\leq 1,$ $ 0\leq \alpha \leq1$  and $A_2$ is a positive constant, within the framework of Lyra's geometry. All types of efforts to have a clear vision of nature of  DE have its own limitations and therefore, still story to find the theoretical aspect behind the accelerated expansion in the universe remains incomplete.

To unveil the mysterious late time cosmic acceleration and DE, now a days, modified gravity theories are more attention seeking and therefore becoming more and more popular for researchers. Among several available modifications of Einstein gravity or general relativity (GR), scalar tensor theories are proved to be the backbone in unlocking various obstacles in clear understanding of the nature of universe such as coincidence problem, cosmic acceleration, inflation and, early and late time behaviors of the universe \cite{bert,bene,saho,chak}. Brans and Dicke proposed a scalar tensor generalization of GR popularly known as Brans Dicke (BD)theory of gravity \cite{bran}, one among the most significant scalar tensor theories because of its wide cosmological inferences \cite{bert,bene} in which  the gravitational interaction includes an extra scalar field $\phi$ other than the metric tensor $g_{ij}$ and $\varpi$ as a dimensionless coupling constant known as BD parameter. It is also to be noticed that this theory reduces to GR if $\phi$ is constant and $\varpi\rightarrow\infty$ \cite{rama1,rama2} but generally it gets possible only if $T$, trace of the energy momentum tensor is non zero.

Some large angle anomalies seen in cosmic microwave back ground (CMB) radiations \cite{spe1,ben} favouring the presence of anisotropies at the early universe thus violate the isotropical nature of the observable universe and hence to clearly describe the early universe, spatially homogeneous but anisotropic Bianchi models play an important role. These Bianchi type models always have attracted attention of researchers working in the area of GR and scalar tensor theories \cite{sing,verm}.
BD theory as a popular scalar tensor theory has been studied by several researchers in different contexts to explore the cosmic evolution of the universe. Johri and Desikan \cite{joh} obtained Brans-Dicke cosmological models with constant deceleration parameter in presence of particle creation. Shri Ram and Singh \cite{shr} studied the effect of time dependent bulk viscosity on the radiation of Friedmann models with zero curvature in BD theory. Singh and Beesham \cite{sing2} studied the effect of bulk viscosity on the evolution of the spatially flat Friedmann-Lemaitre-Robertson-walker models in the context of open thermo-dynamical systems, which allow for particle creation, is analyzed within the framework of BD theory. Reddy et al. \cite{red} discussed homogeneous axially symmetric Bianchi type-I radiating cosmological model with negative constant deceleration parameter in BD scalar-tensor theory of gravitation. Adhav \cite{adh} explored Bianchi type III cosmological model with negative constant deceleration parameter in BD theory of gravity in the presence of perfect fluid. Singh and Sharma \cite{sin4}, Singh \cite{sin5,sin6} studied Bianchi types models in the context of BD scalar-tensor theory of gravitation. Rao and Sudha \cite{rao} investigated Bianchi type V dark energy model in Brans dicke theory of gravitation. Sharif and Waheed \cite{sha1,sha2} investigated Bianchi type I universe models in self interacting BD cosmology.

Motivated by above discussions, here we investigate the spatially homogeneous and totally anisotropic Bianchi type-V cosmological models within the framework of self interacting BD theory of gravity in the background of anisotropic DE with variable EoS parameter and constant deceleration parameter. The EoS parameter $\omega$ and its existing range for the DE models is in good agreement with the most recent observational data. We also determine the most suitable form of $\omega$ as a function of $t$ for the above said models, in explaining the evolution of universe. The physical behaviors of the solutions have been discussed using some physical quantities. The out line of the  work is as follows: In Sec. 2, action functional for self interacting BD theory has been discussed and corresponding field equations for Bianchi type V universe in the presence of anisotropic dark energy fluid have been formulated. The solutions of the field equations have been studied in two different cases, in Sec. 3, we describe the DE model of the universe for $q\neq-1$ and in Sec. 4, we describe the DE model of the universe for $q=-1$. In Sec. 5, we present stability analysis of the models. Finally, we discuss and conclude our work in Sec. 6.
\vskip0.2in
\noindent{\bf 2 Self Interacting Brans Dicke Theory - Action Functional and Field Equations.}
\vskip0.2in
\noindent Brans and Dicke proposed a scalar tensor generalization of GR popularly known as Brans Dicke (BD) theory of gravity \cite{bran}, one of the popular modified gravity theories for researchers working in the area of GR and scalar tensor theory. As a further extension, Sen and Seshadri \cite{sen} defined the action for self interacting Brans Dicke theory (taking G=c=1) as
\begin{equation}\label{1}
S=\frac{1}{16\pi}\int \sqrt{-g} d^4 x\left\{\frac{-\varpi}{\phi}\phi^{,i}\phi_{,i}+\phi R- V(\phi)\right\}+\int L_m \sqrt{-g} d^4 x,
\end{equation}
where $g$ is the determinant of the metric concerned, $\sqrt{-g}\, d^4 x $ denotes four dimensional volume, $\varpi$ is BD coupling constant, $L_m$ denotes the Lagrangian matter part, $\phi$ denotes the scalar field, $V(\phi)$ represents self interacting potential and $R$ represents Ricci scalar. BD field equations can now easily be obtained by varying Eq. \ref{1} with respect to $g_{ij}$ and $\phi$ which are as follows (assuming $\frac{V(\phi)}{2\phi}=\lambda(\phi)$):
\begin{equation}\label{5}
R_{ij}-\frac{1}{2}Rg_{ij}=-\frac{8\pi}{\phi}T_{ij}-\lambda(\phi)g_{ij}-\frac{\varpi}{\phi^2}(\phi_{,i}\phi_{,j}-\frac{1}{2}g_{ij}\phi_{,k}\phi^{,k})-\frac{1}{\phi}(\phi_{i;j}-g_{ij}\Box\phi),
\end{equation}
\begin{equation}\label{6}
\Box\phi=\frac{1}{(3+2\varpi)}\left[2\phi \lambda(\phi)-2\phi^2\frac{d\lambda(\phi)}{d\phi}\right]+\frac{8\pi}{3+2\varpi}T,
\end{equation}
where $T=g^{ij}T_{ij}$ is the trace of the energy-momentum tensor, $\Box$ the d'Alembertian operator, and Comma and semi-colon denote partial and covariant differentiation respectively.\\

\noindent Bianchi type-V spatially homogeneous and anisotropic metric can be defined as following:
\begin{equation}\label{2}
ds^2=dt^2-e^{2A}dx^2-e^{(2B+2\mu x)}dy^2-e^{(2c+2\mu x)}dz^2,
\end{equation}
 where the metric potentials $A$, $B$ and $C$ are functions of cosmic time $t$ and $\mu$ is a parameter. The energy-momentum tensor of anisotropic dark energy fluid is taken as
\begin{eqnarray}\label{3}
T^i_{j}=diag[\rho,-p_x,-p_y,-p_z]
       =diag[1,-\omega_x,-\omega_y,-\omega_z]\rho,
\end{eqnarray}
where $\rho$ is the energy density of the fluid while $p_x$, $p_y$ and $p_z$ are the directional pressures, and $\omega_x$, $\omega_y$ and $\omega_z$ are the directional EoS parameters of the fluid $\omega$ of DE fluid with no deviation along the $x$, $y$ and $z$ axis respectively. Now parameterizing the equation (\ref{3}) by choosing $\omega_z=\omega$ and introducing the skewness parameter $\delta$, which is the deviation of EoS parameter $\omega$ on $x$ and $y$ axis.  Here $\omega$ and $\delta$ are not necessarily constant but the functions of the cosmic time $t$.
\begin{equation}\label{4}
T^i_j=diag[1,-(\omega+\delta),-(\omega+\delta),-\omega]\rho.
\end{equation}
The field equations (\ref{5}) and (\ref{6}) for the metric (\ref{2}) with the help of Eq. (\ref{4}) take the form:
\begin{equation}\label{7}
\dot{A}\dot{B}+\dot{B}\dot{C}+\dot{C}\dot{A}-3\mu^2e^{-2A}=\frac{-8\pi\rho}{\phi}+\lambda(\phi)-\frac{\varpi}{2}\frac{\dot{\phi}^2}{\phi^2}-\frac{1}{\phi}(\ddot{\phi}-\Box\phi)
\end{equation}
\begin{equation}\label{8}
\ddot{B}+\ddot{C}+\dot{B}^2+\dot{C}^2+\dot{B}\dot{C}-\mu^2e^{-2A}=\frac{8\pi\rho(\omega+\delta)}{\phi}-\lambda(\phi)+\frac{\varpi}{2}\frac{\dot{\phi}^2}{\phi^2}-\frac{1}{\phi}(\dot{\phi}\dot{A}-\Box\phi)
\end{equation}
\begin{equation}\label{9}
\ddot{C}+\ddot{A}+\dot{C}^2+\dot{A}^2+\dot{C}\dot{A}-\mu^2e^{-2A}=\frac{8\pi\rho(\omega+\delta)}{\phi}-\lambda(\phi)+\frac{\varpi}{2}\frac{\dot{\phi}^2}{\phi^2}-\frac{1}{\phi}(\dot{\phi}\dot{B}-\Box\phi)
\end{equation}
\begin{equation}\label{10}
\ddot{A}+\ddot{B}+\dot{A}^2+\dot{B}^2+\dot{A}\dot{B}-\mu^2e^{-2A}=\frac{8\pi(\rho\omega)}{\phi}-\lambda(\phi)+\frac{\varpi}{2}\frac{\dot{\phi}^2}{\phi^2}-\frac{1}{\phi}(\dot{\phi}\dot{C}-\Box\phi)
\end{equation}
\begin{equation}\label{11}
\dot{B}+\dot{C}-2\dot{A}=0
\end{equation}
\begin{equation}\label{12}
\Box\phi=\ddot{\phi}+\dot{\phi}(\dot{A}+\dot{B}+\dot{C})=\frac{1}{3+2\varpi}\left[8\pi(\rho-3p)-2\phi^2\frac{d\lambda(\phi)}{d\phi}+2\phi \lambda(\phi)\right]
\end{equation}
\noindent The energy conservation equation $T^{ij}_{;j}=0$ takes the form as:
\begin{equation}\label{13}
\dot{\rho}+3(1+\omega)\rho H+\rho(\delta H_x)+\rho(\delta H_y)=0,
\end{equation}
where  the over dot $(\cdot)$ denotes derivative with respect to the cosmic time $t$.
The directional Hubble parameters that express the expansion rates of the universe along the direction of $x$, $y$ and $z$ axis can be defined as:
\begin{equation}\label{14}
H_x=\dot{A},  H_y= \dot{B},  H_z=\dot{C}.
\end{equation}
The generalized mean Hubble parameter $H$, which expresses the expansion rate of the universe, is given by
\begin{equation}\label{15}
H=\frac{1}{3}\left(H_x+H_y+H_z\right).
\end{equation}
The spatial volume $V$ is given as
\begin{equation}\label{16}
V^3=e^{(A+B+C+2\mu x)}.
\end{equation}
The mean anisotropic parameter of the expansion $\Delta$ has a very crucial role in deciding whether the model is isotropic or anisotropic and is defined as:
\begin{equation}\label{17}
\Delta=\frac{1}{3}\frac{\sigma^2}{H^2}=\frac{1}{3}\sum_{i=x}^z\left(\frac{H_i-H}{H}\right)^2,
\end{equation}
where $\sigma$ is the shear scalar, $\Delta$ is the measure of the deviation from isotropic expansion, the universe expands isotropically when $\Delta=0$. The model approaches to isotropy continuously if $V\rightarrow\infty$ and $\Delta=0$ as $t\rightarrow\infty$ \cite{col}. For more physical realistic approach of the models, we expect the energy densities of the dark energy fluid to be positive as $t\rightarrow\infty$.
Let us now find the expansion and shear scalar for the metric (\ref{2}). The expansion scalar is given by
\begin{equation}\label{18}
\Theta=3H,
\end{equation}
and the shear scalar is given by
\begin{equation}\label{19}
\sigma^2=\frac{1}{2}\left[\sum_{i=x}^zH_i^2-3H^2\right].
\end{equation}
The dimensionless mean deceleration parameter (DP) $q$ in cosmology is the measure of the cosmic acceleration of the universe expansion. The universe exhibits accelerating volumetric expansion when $-1\leq q<0$,  decelerating volumetric expansion when $q>0$ and constant  when $q=0$. It is mentioned here that $q$ was supposed to be positive initially but recent observations from the supernova experiments suggest that it is negative. Thus the behavior of the universe models depends upon the sign of $q$. The positive deceleration parameter corresponds to a decelerating model while the negative value provides inflation.
We assume that the deceleration parameter $q$ is constant \cite{ber1}
\begin{equation}\label{20}
q=-\frac{V\ddot{V}}{\dot{V}^2}=constant.
\end{equation}
\noindent Solving the equation (\ref{20}), we get
\begin{equation}\label{21}
V=(q_0 t-q_1)^{\frac{1}{(1+q)}}, \,\,q\neq-1
\end{equation}
\begin{equation}\label{22}
V= e^{(q_2t+q_3)},\,\, q=-1
\end{equation}
where $q_0,q_1,q_2$ and $q_3$ are arbitrary constants. Thus we obtain two values of the spatial volume which correspond to two different models of the universe i.e. power law model and exponential model of the universe.
\vskip0.2in
\noindent{\bf 3  DE power law model of the universe when $q\neq-1$.}
\vskip0.2in
\noindent Now we discuss the model of universe when $q\neq-1$, i.e., $V=(q_0 t-q_1)^{\frac{1}{(1+q)}}$. To get the deterministic model, we assume \cite{sin5}
\begin{equation}\label{23}
B=mA,
\end{equation}
where $m$ is a positive constant. Now solving the field Eq. (\ref{11}) with the help of Eqs. (\ref{16}), (\ref{21}) and (\ref{23}), we obtain the expressions for metric coefficients as
\begin{equation}\label{24}
A=\frac{1}{(1+q)}\ln(q_0t-q_1)-\frac{2}{3}\mu x,
\end{equation}
\begin{equation}\label{25}
C=\frac{(2-m)}{(1+q)}\ln(q_0t-q_1)-\frac{2}{3}\mu x.
\end{equation}
From above equations, it can be observed that in this cosmological model, universe exhibits initial singularity of the POINT-type at $t=\frac{q_1}{q_0}$ and otherwise space time is well behaved.
Subtracting Eq (\ref{8}) from (\ref{9}), and using Eqs. (\ref{23})-(\ref{25}), we get
\begin{equation}\label{26}
\phi(t)=\phi_1\frac{(q_0t-q_1)^{\phi_2}}{\exp(2\mu x)},
\end{equation}
where $\phi_0$ is an arbitrary integration constant, $\phi_1=\frac{\phi_0 q_0}{(1+q)}$, $\phi_2=\frac{(2-q)}{(1+q)}$ and other physical quantities $\rho$, $\lambda(\phi)$ and $\delta$ are obtained as
\begin{eqnarray}\label{27}
\rho(t)&=&\phi_1\frac{(q_0t-q_1)^{\phi_2}}{8\pi (\omega-1) \exp(2\mu x)}\{\rho_1+\rho_2+\rho_3\},
\end{eqnarray}
where $$\rho_1=\frac{3q_0^2(m+1)}{(1+q)^2 (q_0t-q_1)^2}-\frac{q_0^2(m+1)}{(1+q)(q_0t-q_1)^2} ,$$
$$\rho_2=-\frac{4\mu^2 \exp(\frac{4}{3}\mu x)}{(q_0t-q_1)^{\frac{2}{(1+q)}}}+\frac{(2-m)\phi_2 q_0^2}{(1+q)(q_0t-q_1)^2} ,$$
$$\rho_3=\frac{\phi_2 q_0^2 (\phi_2-1)}{(q_0t-q_1)^2}-\frac{2\phi_2 q_0^2}{(q_0t-q_1)^2}\left((\phi_2-1)+\frac{3}{1+q}\right).$$
\begin{eqnarray}\label{28}
\lambda(\phi)\simeq\lambda(t)=\frac{\lbrace\lambda_1+\lambda_2+\lambda_3\rbrace}{2(q_0 t-q_1)^2 (1+q)^2 (\omega-1)}- \frac{\mu^2\exp(\frac{4}{3}\mu x)(1+3\omega)}{(\omega-1)(q_0t-q_1)^{\frac{2}{(1+q)}}},
\end{eqnarray}
where $$\lambda_1=-2q_0^2(m^2-2m-2)(\omega-1)+\varpi(\phi_2q_0)^2(1+q)^2(\omega-1),$$
$$\lambda_2=-6(1+q)\phi_2q_0^2(\omega+1)+2q_0^2(m+1)(2-q),$$
$$\lambda_3=2(1+q)\phi_2q_0^2\left\{(2-m)-(1+q)(\phi_2-1)\right\}.$$
\begin{eqnarray}\label{29}
\delta(t)=\frac{\phi_1q_0^2(1-m)\left\{3-(1+q)(1+\phi_2)\right\}}{\exp(2\mu x)4\pi \rho (1+q)^2(q_0 t-q_1)^{2-\phi_2}},
\end{eqnarray}

\begin{figure}[htbp]
\centering
\includegraphics[width=2.0in]{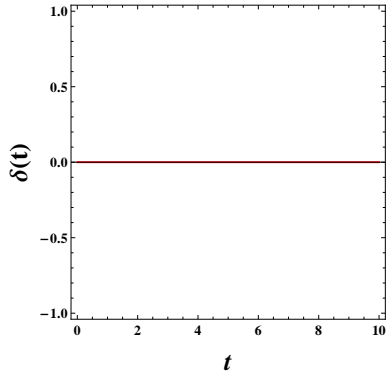}
\caption{The plot of the skewness parameter ($\delta$) vs. cosmic time ($t$).}
\label{fig:fig1}
\end{figure}
\noindent From Fig. 1, it is clear that in self interacting BD theory, in case of Bianchi type V power law model, hardly any anisotropy is seen in DE fluid.

\noindent The physical parameters, \textit{viz.}, the mean generalized Hubble parameter $H$, the mean anisotropy parameter $\Delta$, the expansion scalar $\Theta$, shear scalar $\sigma^2$, and the ratio $\frac{\sigma}{\Theta}$ are given by
\begin{equation}\label{31}
H=\frac{q_0}{(1+q)(q_0t-q_1)},
\end{equation}
\begin{equation}\label{32}
\Delta=\frac{(m-1)^2}{3},
\end{equation}
\begin{equation}\label{33}
\Theta=\frac{3q_0}{(q+1)(q_0 t-q_1)},
\end{equation}
\begin{equation}\label{34}
\sigma^2=\frac{q_0^2(m-1)^2}{(q+1)^2(q_0t-q_1)^2},
\end{equation}
\begin{equation}\label{35}
\frac{\sigma}{\Theta}=\frac{m-1}{3}.
\end{equation}

\noindent In this cosmological model, we observe that the universe possesses initial singularity of the POINT- type at $t=\frac{q_1}{q_0}$. The space-time is well behaved in the range $\frac{q_1}{q_0}<t<\infty$. At the initial moment $t=\frac{q_1}{q_0}$, the parameters $H$, $\Theta$, and $\sigma^2$ diverge. So the universe starts from initial singularity with infinite rate of shear and expansion. Moreover $H$, $\Theta$, and $\sigma^2$ tend to zero as $t\rightarrow\infty$. Therefore the expansion stops and shear dies out. Here $H$, $\Theta$, and $\sigma^2$ are monotonically decreasing toward a zero quantity for $t$ in the range $\frac{q_1}{q_0}<t<\infty$. The proper volume $V$ tends to zero at the initial singularity. As time proceeds the universe approaches toward an infinitely large volume in the limit as $t\rightarrow\infty$.  The model describes an accelerating model when $q<0$ and a decelerating model parameter for $q>0$. The ratio $\frac{\sigma^2}{\Theta^2}$ of the model is non zero constant for $m\neq1$. This shows that the model does not approach to isotropy at the time of the evolution of the universe. The mean anisotropic parameter $\Delta$ is uniform throughout the evolution of the universe and hence we can say that model discussed above remains anisotropic throughout for $m\neq1$

In nutshell, we can say that above model starts with zero volume and heads towards infinitely large volume remaining anisotropic throughout.
\vskip0.1in
\noindent\textit{3.1  DE power law model of the universe when EoS parameter $\omega$ is constant}
\vskip0.1in
\noindent Equation (\ref{27}) depicts that $\rho$ i.e. energy density depends on EoS parameter $\omega$. Therefore, now we explore $\rho$ assuming constant values of $\omega$ describing various forms of universe depending upon the value of $\omega$ e.g. the values of $\omega= -1,0,1/3$ represent vacuum dominated, dust dominated and radiation dominated era respectively. Nature of $\rho$ has been shown in Fig. 2. We observe that in all three cases, energy density in the universe decreases with time with different rates. Also, the nature of self interacting potential $V(\phi)$ with $\phi$ has been clearly shown in Fig. 3.
\begin{figure}[htbp]
\centering
\includegraphics[width=2.0in]{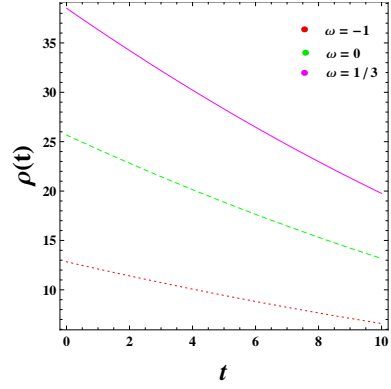}
\caption{The plot of energy density ($\rho$) vs. cosmic time ($t$).}
\label{fig:fig1}
\end{figure}
\begin{figure}[htbp]
\centering
\includegraphics[width=2.2 in]{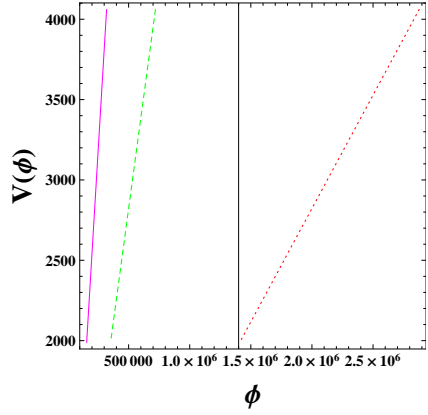}
\caption{The plot of the self interacting potential $\lambda(\phi)$ vs. scalar field ($\phi$).}
\label{fig:fig2}
\end{figure}
\vskip0.1in
\noindent\textit{3.2  DE power law model of the universe when EoS parameter $\omega$ is variable}
\vskip0.1in
\noindent Now, we study the nature of $\rho$ assuming $\omega$ as variable. Without loss of generality, we discuss three forms of $\omega(t)$ which can also be constrained using observational data, as:
\begin{equation}\label{30a}
\omega(t)=k_1t^2+k_2t+k_3,
\end{equation}
\begin{equation}\label{30b}
\omega(t)=\frac{-t e^{k_1t}}{10^5},
\end{equation}
\begin{equation}\label{30c}
\omega(t)=log(k_1t).
\end{equation}
Evolution of above mentioned forms of  $\omega(t)$ has been shown in Fig. 3 as red dotted curve, green dashed curve, and the magenta solid curve corresponding to $\omega(t)=k_1t^2+k_2t+k_3$, $\omega(t)=\frac{-t e^{k_1t}}{10^5}$ and $\omega(t)=log(k_1t)$ respectively. As we see that $\omega$ is a function of cosmic time $t$. Therefore it can be either a function of redsift $z$ or the scale factor $R$. There are several parametrizations of $\omega(z)$ that can be either linearly as
\begin{equation}\label{42}
\omega(z) = \omega_0+\omega'z,
\end{equation}
where $\omega'=\left[\frac{d\omega}{dz}\right]_{z=0}$ \cite{hut} or nonlinearly as
\begin{equation}\label{43}
\omega(z) = \omega_0+\frac{\omega'z}{1+z},
\end{equation}
\cite{chev}. The parametrization of scale factor dependence of $\omega$ is given as
\begin{equation}\label{44}
\omega(R) = \omega_0+\omega_R(1-R),
\end{equation}
where $\omega_0$ is the present value at $R=1$ and $\omega_R$ is the measure of the time variation $\omega'$ \cite{lind}. If we compare our assumptions with the experimental results then we conclude that the limit of $\omega$ accommodates well within the acceptable range of EoS parameter as described in first para of Sec. 1.  For this model, the values of EoS parameter $\omega$ can be restricted to the observational limits \cite{kom1}. From Fig. 4, we also observe that each curve passes through the various phases of DE model of the universe with appropriate choices of integrating constant.
\begin{itemize}
\item Phantom phase when $\omega<-1$,
\item Quintum  phase, which inherits both the properties of quintessence and phantom phase by the  phantom divide line  $\omega=-1$,
\item the cosmological constant $\Lambda$, which is considered to be the simplest case of the DE when $\omega=-1$,
\item Quintessence phase when $-1<\omega<\frac{-1}{3}$ \cite{car3},
\end{itemize}

Thus from Fig. 4, we say that the earlier DE dominated phase of the universe transited into the matter dominated phase of the universe in case of curves (\ref{30a}) and (\ref{30c}), where as the earlier matter dominated phase of the universe  transited into the DE dominated phase of the universe in case of curve (\ref{30b}) at late times.
\begin{figure}[htbp]
\centering
\includegraphics[width=2.0 in]{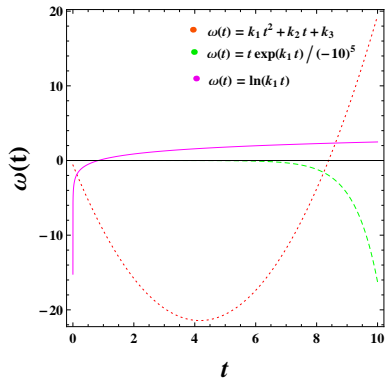}
\caption{The plot of EoS parameter ($\omega$) vs. cosmic time ($t$).}
\label{fig:fig1}
\end{figure}
\begin{figure}[htbp]
\centering
\includegraphics[width=2.0in]{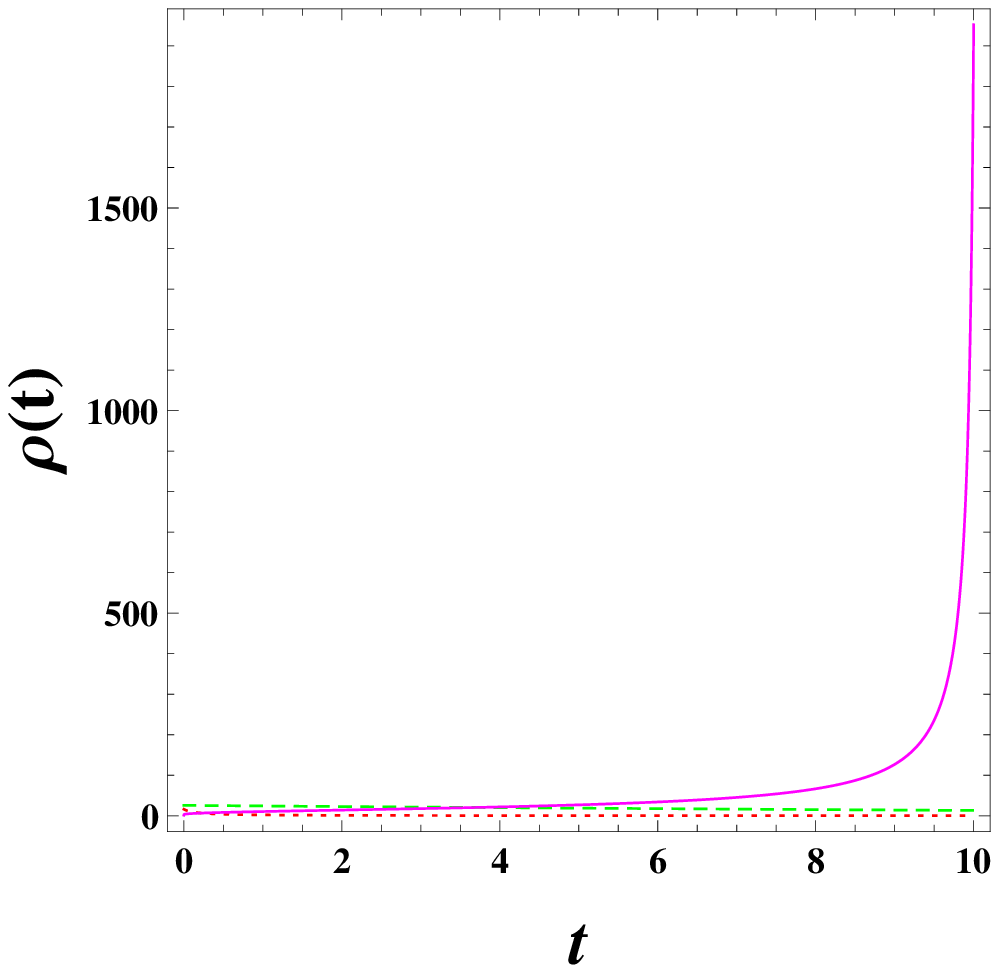}
\caption{The plot of energy density ($\rho$) vs. cosmic time ($t$).}
\label{fig:fig2}
\end{figure}

\begin{figure}[htbp]
\centering
\includegraphics[width=2.0 in]{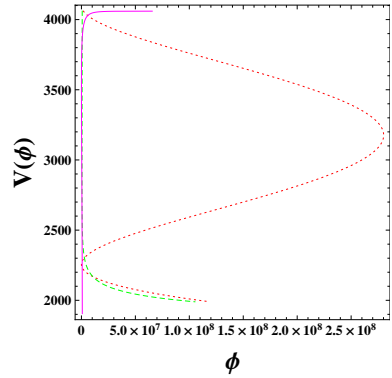}
\caption{The plot of the self interacting potential $V(\phi)$ vs. scalar field ($\phi$).}
\label{fig:fig1}
\end{figure}

Nature of energy density with time $t$ and self interacting potential with scalar field $\phi$ respectively has been shown as red dotted curve, green dashed curve, and the magenta solid curve corresponding to $\omega(t)$ given by (\ref{30a}), (\ref{30b}), (\ref{30c}) respectively in Figs. 5 and 6.
\vskip0.2in
\noindent{\bf 4  DE exponential model of the universe when $q=-1$}
\vskip0.2in
\noindent In this section, we discuss the model of universe when $q=-1$, i.e., $ V=e^{(q_2t+q_3)}$. Solving the field Eq. (\ref{11}) with the help of Eqs. (\ref{16}), (\ref{22}) and (\ref{23}), we obtain the expressions for metric coefficients as
\begin{equation}\label{36}
A=(q_2t+q_3)-\frac{2}{3}\mu x,
\end{equation}
\begin{equation}\label{37}
C=(2-m)\left\{(q_2t+q_3)-\frac{2}{3}\mu x\right\}.
\end{equation}
Subtracting Eq. (\ref{8}) from (\ref{9}), and using Eqs. (\ref{23}), (\ref{36}) and (\ref{37}), we get
\begin{equation}\label{38}
\phi(t)=\phi_0 q_2\exp\left\{3(q_2t+q_3)-2\mu x)\right\},
\end{equation}
The physical quantities $\rho$, $\lambda(\phi)$ and $\delta$ are obtained as:
\begin{eqnarray}\label{39}
\rho(t)&=& \frac{\phi_0 q_2}{8\pi (1-\omega)}\rho_1^*(\rho_2^*+\rho_3^*),
\end{eqnarray}
where $$\rho_1^*=\exp\left\{3(q_2t+q_3)-2\mu x\right\},$$
$$\rho_2^*=18q_2^2,$$
$$\rho_3=4\mu^2\exp\left\{\frac{4}{3}\mu x-2(q_2 t+q_3)\right\}.$$
\begin{eqnarray}\label{40}
\lambda(\phi)\simeq\lambda(t)=\lbrace\lambda_1^*-\lambda_2^*+\lambda_3^*\rbrace,
\end{eqnarray}
where $$\lambda_1^*=q_2^2\left\{9\left(\frac{\varpi}{2}+1-2q_2^2\right)-(m^2-2m-2)\right\},$$
$$\lambda_2^*=3\mu^2\exp\left\{\frac{4}{3}\mu x-2(q_2t+q_3)\right\},$$
$$\lambda_3^*=\frac{18q_2^2+4\mu^2\exp\left\{\frac{4}{3}\mu x-2(q_2 t+q_3)\right\}}{1-\omega}.$$
\begin{eqnarray}\label{41}
\delta(t)=0.
\end{eqnarray}
The physical parameters, \textit{viz.}, the mean generalized Hubble parameter $H$, the mean anisotropy parameter $\Delta$, the expansion scalar $\Theta$, shear scalar $\sigma^2$, and the ratio $\frac{\sigma}{\Theta}$ are given by
\begin{equation}\label{43}
H=q_2,
\end{equation}
\begin{equation}\label{44}
\Delta=\frac{(m-1)^2}{3},
\end{equation}
\begin{equation}\label{45}
\Theta=3q_2,
\end{equation}
\begin{equation}\label{46}
\sigma^2=q_2^2(m-1)^2,
\end{equation}
\begin{equation}\label{47}
\frac{\sigma}{\Theta}=\frac{m-1}{3}.
\end{equation}
\noindent In this case, we observe that it is an exponential model of the universe and hence this model is non-singular because the exponential function is never zero. The physical parameters $H$, $\Delta$, $\Theta$ and $\sigma^2$ are all finite and uniform throughout the evolution of the universe. The volume scale factor $V$ increases exponentially with time for $q_2, q_3>0$, which indicates that the universe starts its expansion with zero volume from infinite past. The expansion scalar $\Theta$ is always positive for $q_2>0$ and the decelerating parameter $q$ is $-1$. Therefore the DE model describes an accelerating  expansion of the universe. Here also, the ratio $\frac{\sigma^2}{\Theta^2}$ of the model is non zero constant for $m\neq1$. The mean anisotropic parameter $\Delta$ is uniform throughout the evolution of the universe as it does not depend on $t$. In this model, the skewness parameter $\delta$ vanishes. Thus, we observe that there is no deviation of EoS parameter $\omega$ on $x$ and $y$ axis and the anisotropic DE fluid disappears in this model.
\vskip0.1in
\noindent\textit{4.1  DE exponential model of the universe when EoS parameter $\omega$ is constant}
\vskip0.1in
\begin{figure}[htbp]
\centering
\includegraphics[width=2.0 in]{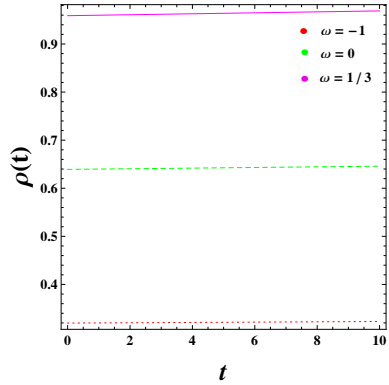}
\caption{The plot of energy density ($\rho$) vs. cosmic time ($t$).}
\label{fig:fig1}
\end{figure}
\begin{figure}[htbp]
\centering
\includegraphics[width=2.0in]{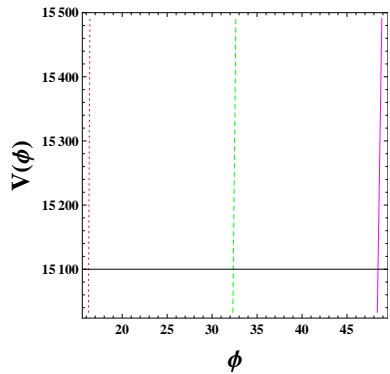}
\caption{The plot of the self interacting potential $V(\phi)$ vs. scalar field ($\phi$).}
\label{fig:fig2}
\end{figure}
\noindent Behaviour of energy density $\rho$ with time $ t $ and self interacting potential $V(\phi)$ with scalar field $\phi$ in case of constant EoS parameter $\omega$ can easily be understood by the Figs. 7 and 8 with suitable choices of integrating constants.
\vskip0.1in
\noindent\textit{4.2  DE exponential model of the universe when EoS parameter $\omega$ is variable}
\vskip0.1in
\begin{figure}[htbp]
\centering
\includegraphics[width=2.0 in]{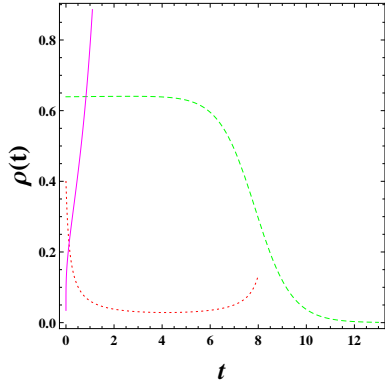}
\caption{The plot of energy density ($\rho$) vs. cosmic time ($t$).}
\label{fig:fig1}
\end{figure}
\noindent Here, we discuss the evolution of energy density and self interacting potential with variable EoS parameter. Without loss of generality we take the EoS parameter $\omega$  as given by (\ref{30a}), (\ref{30b}) and (\ref{30c}). Corresponding behaviour of $\rho$ and $V(\phi)$ has been shown in Figs. 9 and 10.
\begin{figure}[htbp]
\centering
\includegraphics[width=2.0 in]{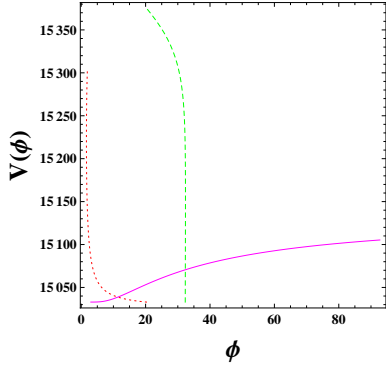}
\caption{The plot of the self interacting potential $V(\phi)$ vs. scalar field ($\phi$).}
\label{fig:fig2}
\end{figure}
\vskip0.2in
\noindent{\bf 5 Stability Analysis}
\vskip0.2in
\begin{figure*}
\begin{center}
$\begin{array}{c@{\hspace{.1in}}c c}
\epsfxsize=1.6in
\epsffile{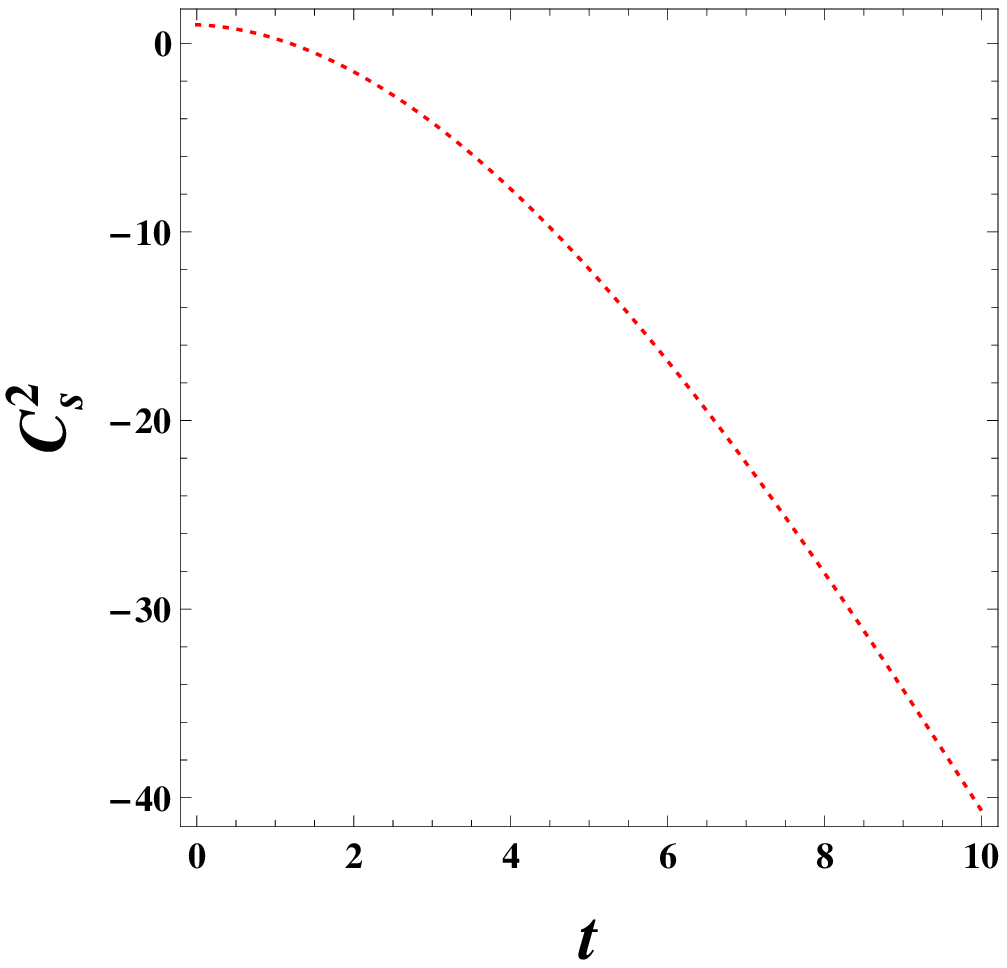} &
\epsfxsize=1.6in
	\epsffile{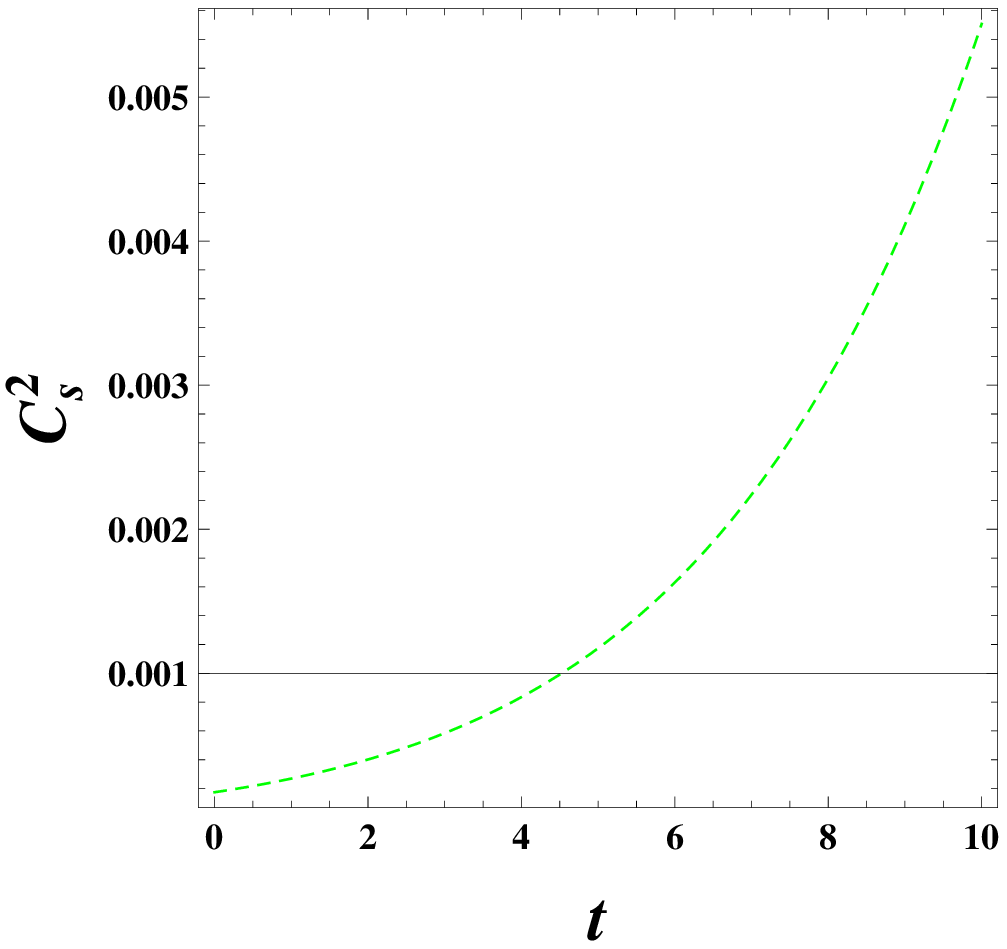} &
	\epsfxsize=1.5in
	\epsffile{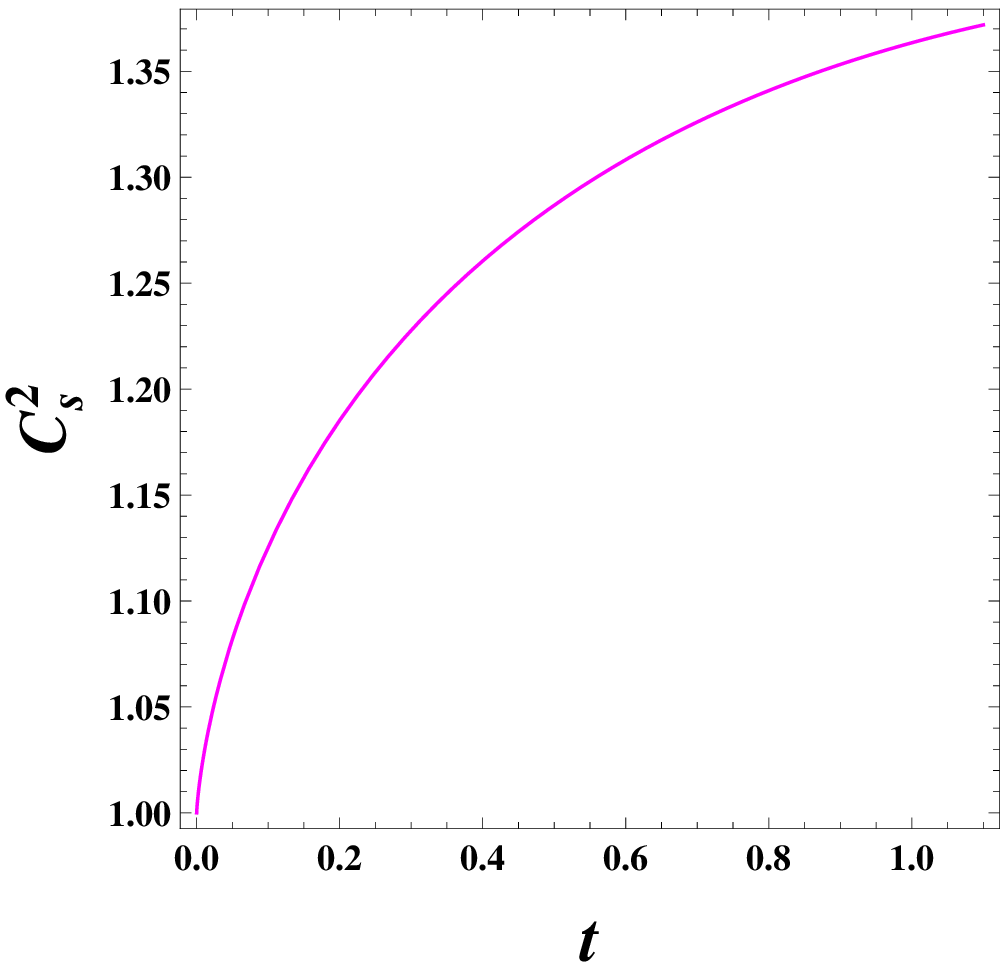} \\
\mbox{\bf (i)} & \mbox{\bf (ii)} & \mbox{\bf (iii)}
\end{array}$
$\begin{array}{c@{\hspace{.1in}}c c}
\epsfxsize=1.6in
\epsffile{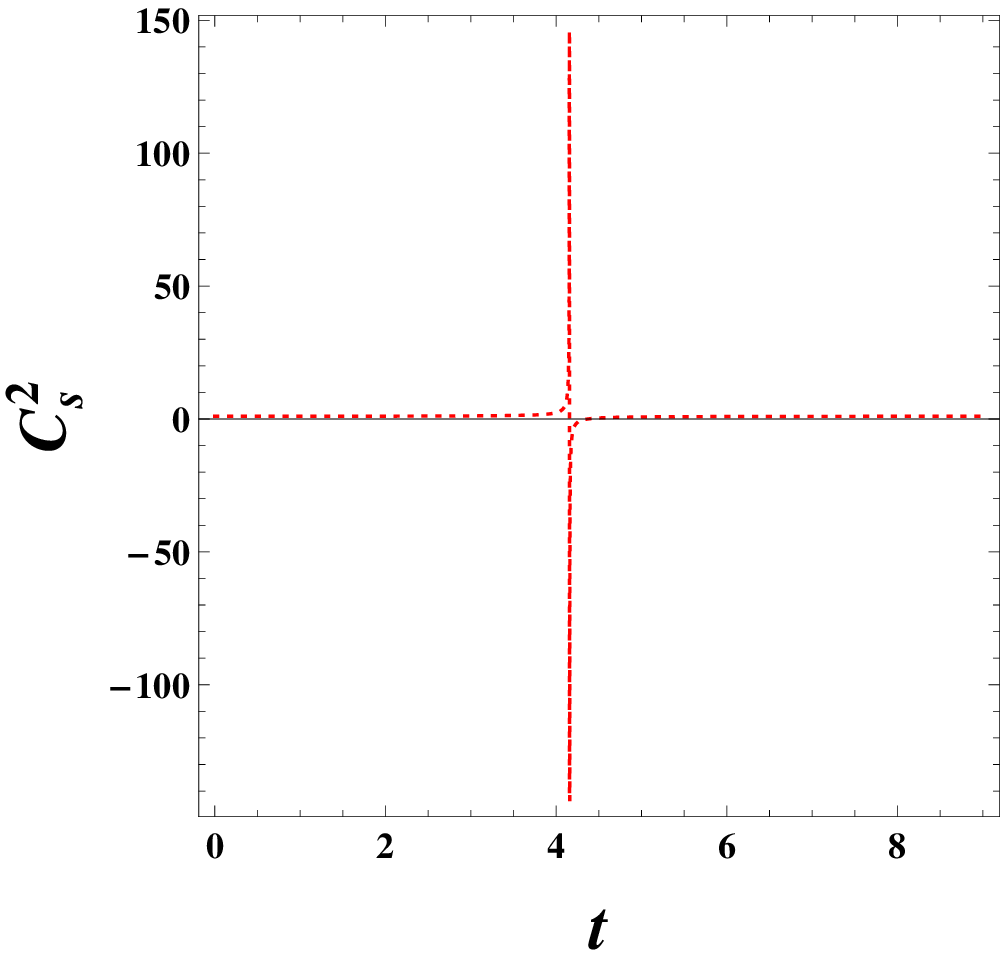} &
	\epsfxsize=1.6in
	\epsffile{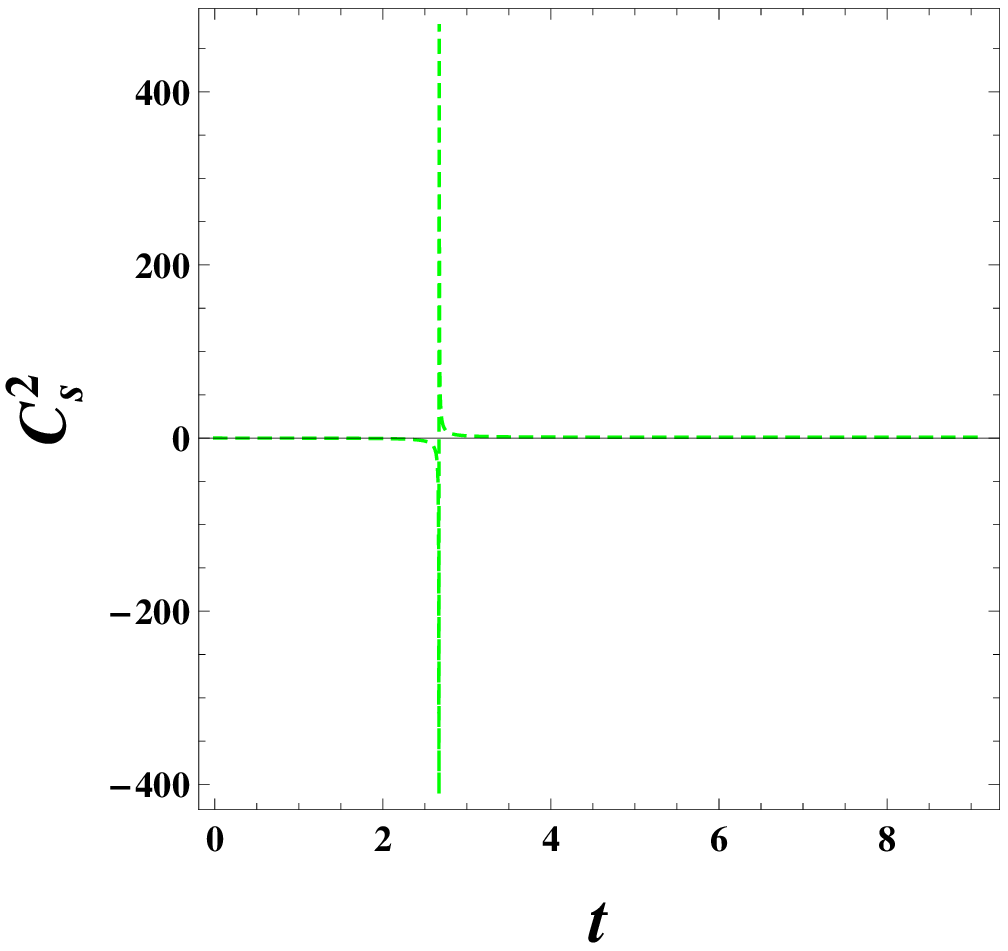} &
	\epsfxsize=1.6in
	\epsffile{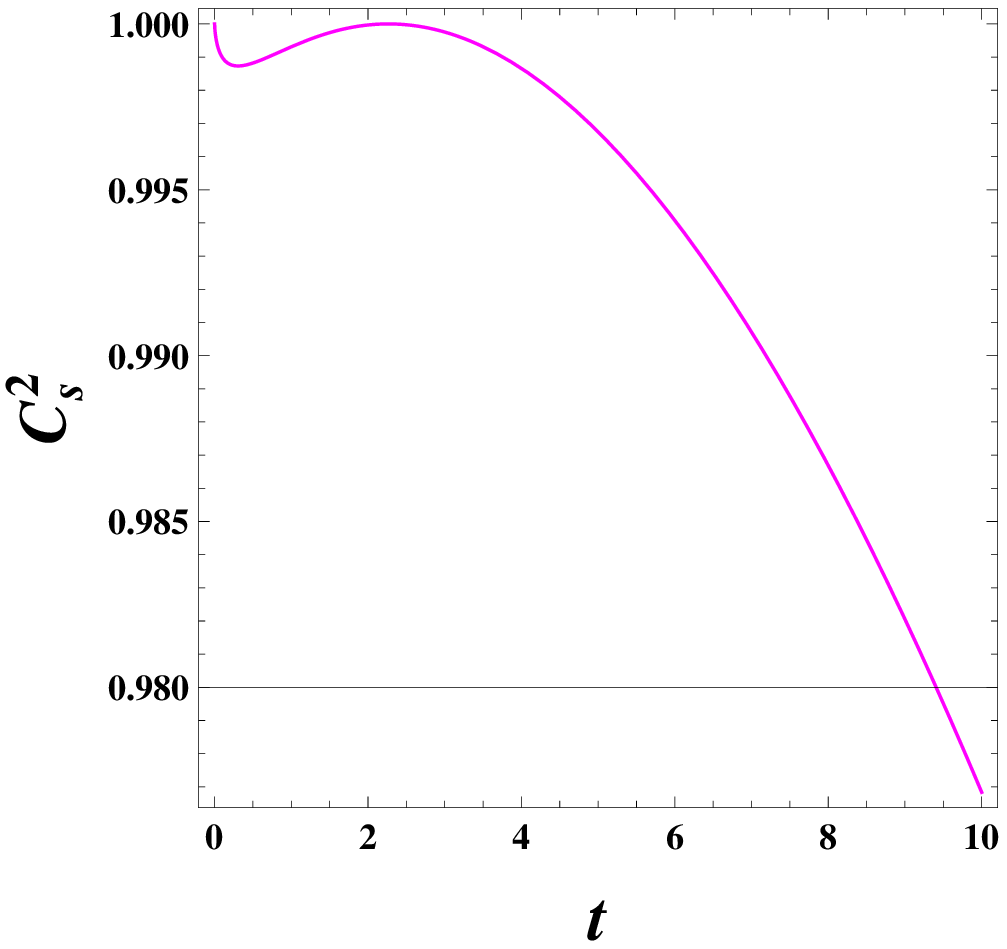} \\
\mbox{\bf (iv)} & \mbox{\bf (v)} & \mbox{\bf (vi)}
\end{array}$
\end{center}
\caption{ This figure corresponds to the sound speed ($C_s^2$) in terms of time $t$ when equation of state $\omega$ is variable. Panels (i), (ii) and (iii) represent $C_s^2$ for DE model of the universe when $q\neq-1$, and   Panels (iv), (v) and (vi) represent $C_s^2$ for DE model of the universe when $q=-1$.}
\label{hsn}
\end{figure*}
\noindent There are various methods available to investigate stability of the models. In the present analysis, we use sound speed given as $$C_s^2=\frac{dp}{d\rho}.$$
For a model to be physically acceptable, $C_s^2\geq 0$. Respective plots of $C_s^2$ in terms of $ t $ for both of the models discussed in Secs. 3 and 4 for variable EoS parameter $\omega$ given by (\ref{30a}), (\ref{30b}) and (\ref{30c}) has been shown in Fig. 11 from panels (i)-(vi). We examine that the DE models of universe within the framework of self interacting BD theory of gravity are stable in some particular cosmic time frames but not throughout the evolution of the universe. Also by comparing the plots of $C_s^2$ for all three assumed EoS parameter for both of the models, we can say that $\omega$ given by (\ref{30c}) is more suitable in explaining the evolution of the universe.
\vskip0.2in
\noindent{\bf 6 Conclusions}
\vskip0.2in
\noindent This paper is devoted to explore the possible solutions of Bianchi type-V cosmological models within the framework of self interacting BD theory of gravity in the background of anisotropic dark energy with the assumption of constant deceleration parameter that leads to two models of the universe, i.e. power law model and exponential model. These models have been discussed in two different cases one with constant EoS parameter $ \omega $, and other with variable EoS parameter $ \omega $. We observed that in self interacting BD theory, anisotropy in DE fluid does not exist in both of i.e. Bianchi type V, power law and exponential universe models respectively. Also we observed that the variable EoS parameter $\omega$ given by (\ref{30a}), (\ref{30b}) and (\ref{30c}) and its existing range is in good agreement with the most recent observational data \cite{kom1}. Comparing the plots of $C_s^2$ for all three assumed EoS parameter for both of the models, we noticed that $\omega$ given by (\ref{30c}) i.e $\omega(t)=log(k_1t)$ is more suitable in explaining the evolution of the universe.

\begin{itemize}
\item In power law model of the universe, we observed that the universe exhibits initial singularity of the POINT- type at $t=\frac{q_1}{q_0}$. The space-time is well behaved in the range $\frac{q_1}{q_0}<t<\infty$. At the initial moment $t=\frac{q_1}{q_0}$, the parameters $H$, $\Theta$, and $\sigma^2$ diverge. So the universe starts from initial singularity with infinite rate of shear and expansion. Moreover $H$, $\Theta$, and $\sigma^2$ tend to zero as $t\rightarrow\infty$. Therefore the expansion stops and shear dies out. Here $H$, $\Theta$, and $\sigma^2$ are monotonically decreasing toward a zero quantity for $t$ in the range $\frac{q_1}{q_0}<t<\infty$. The shear tends to zero much faster than the expansion. The proper volume $V$ tends to zero at the initial singularity. As time proceeds the universe approaches toward an infinitely large volume in the limit as $t\rightarrow\infty$.  The ratio $\frac{\sigma^2}{\Theta^2}$ of the model is non zero constant for $m\neq1$. This shows that the model does not approach to isotropy at the time of the evolution of the universe.

\item The exponential model of the universe corresponds to $q=-1$ with spatial volume as  $ V=e^{(q_2t+q_3)}$. The physical singularity does not exist in this model because the exponential function is asymptotic at the infinite past. The physical parameters  $H$, $\Delta$, $\Theta$, and $\sigma^2$ are all finite as the metric functions do not vanish for this model. The volume scale factor $V$ increases exponentially with time, which indicates that the universe starts its expansion with zero volume from infinite past. The expansion scalar $\Theta$ is positive for $q_2>0$. Therefore, the DE model describes an accelerating  expansion of the universe in this case.

\item In Sec. 5, we examine the stability of power law model as well as exponential model. In both the cases, detecting the values of $ C_s^2$ from the respective plots, we observe that models are stable but not throughout the evolution of the universe.
\end{itemize}

Both of the above anisotropic models  describe accelerating expansion of the universe but fail to provide time or red-shift based transition of the universe from deceleration to acceleration, because of the constant value of the deceleration parameter $q$. Thus from the above mentioned results, one can clearly depict that despite having several prominent features, the above discussed models fail in details.


\noindent\begin{thebibliography}{999999}
\bibitem{kom1} Komatsu, E., et al.: Astrophys. J. Suppl. Ser. \textbf{192}, 18 (2011).
\bibitem{per3} Perlmutter, S., et al.: Astrophys. J. \textbf{565}, 517 (1999).
\bibitem{rie1} Riess, A. G., et al.: Astron. J. \textbf{116}, 1009 (1998).
\bibitem{spe1} Spergel, D.N., et al.: ApJS \textbf{148}, 175 (2003).
\bibitem{spe2} Spergel, D.N., et al.: ApJS \textbf{170}, 377 (2007).
\bibitem{teg} Tegmark, M., et al. (SDSS Collboration): Phys. Rev. D \textbf{69}, 103501 (2004).
\bibitem{ben} Bennett, C. L., et al.: Astrophys. J. Suppl. Ser. \textbf{148}, 1 (2003).
\bibitem{all} Allen, S. W., et al.: Mon. Not. R. Astron. Soc. \textbf{353}, 457 (2004).
\bibitem{jai} Jain, B., Taylor, A.: Phys. Rev. Lett. \textbf{91}, 141302 (2003).
\bibitem{pla} Ade, P. A. R. et al. [Planck Collaboration], arXiv:1303.5076 [astro-ph.CO].
\bibitem{car3} Carroll, S. M., Hoffman, M.: : Phys. Rev. D \textbf{68}, 023509 (2003).
\bibitem{kuj} Kujat, J., et al.: Astrophys. J. \textbf{572}, 1 (2002).
\bibitem{bar} Bartelmann, M., et al.: New Astron. Rev. \textbf{49}, 199 (2005).
\bibitem{jim} Jimenez, R.: New Astron. Rev. \textbf{47}, 761 (2003).
\bibitem{das} Das, A., et al.: : Phys. Rev. D \textbf{72}, 043528 (2005).
\bibitem{rat} Ratra, B., Peebles, P. J. E.: Phys. Rev. D \textbf{37}, 3406 (1988).
\bibitem{sah2} Sahni, V., Starobinsky, A. : Int. J. Mod. Phys. D \textbf{15}, 2105 (2006).
\bibitem{sah3} Sahni, V., Shafielooa, A., Starobinsky, A. : Phys. Rev. D \textbf{78}, 103502 (2008).
\bibitem{rah} Rahaman, F., Bhui, B., Bhui, B. C.: Astrophys. Space Sci. \textbf{301}, 47 (2006).
\bibitem{muk} Mukhopadhyay, U., Gosh, P. P., Choudhury, S. B. D.: Int. J. Mod. Phys. D \textbf{17}, 301 (2008).
\bibitem{bam} Bamba, K., Capozziello, S., Nojiri, S. and Odintsov, S.D.: Astrophys. Space Sci. arXiv:1205.3421v3 \textbf{342}, 155 (2012).
\bibitem{aka} Akarsu, \"{O}., Kilinc, C. B.: Gen. Rel. Grav. \textbf{42}, 763 (2010).
\bibitem{sin1} Singh, J. K., Sharma, N. K.: Int. J. Theor. Phys. \textbf{53}, 1375 (2014).
\bibitem{sin2} Singh, J. K., Sharma, N. K.: Int. J. Theor. Phys. \textbf{53}, 1424 (2014).
\bibitem{sin3} Singh, J. K., Sharma, N. K.: Int. J. Theor. Phys. \textbf{53}, 461 (2014).
\bibitem{sir2} Singh, J. K., Rani, S.: Appl. Math. comput. \textbf{259}, 187 (2015).
\bibitem{bert}Bertolami, O., Martins, P.J.: Phys. Rev. D \textbf{61}, 064007 (2000).
\bibitem{bene}Benerjee, N., Pavon, D.: Phys. Rev. D \textbf{63}, 043504 (2001).
\bibitem{saho}Sahoo, B. K. and Singh, L. P.: Mod. Phys. Lett. A \textbf{18}, 2725 (2003).
\bibitem{chak} Chakraborty, W. and Debnath U.: Int. J. Theor. Phys. \textbf{48}, 232 (2009).
\bibitem{bran} Brans, C., Dicke, R. H.: Phys. Rev. \textbf{124}, 925 (1961).
\bibitem{rama1} Rama, S. K., and Gosh, S.: Phys. Lett. B \textbf{383}, 32 (1996)..
\bibitem{rama2} Rama, S. K.: Phys. Lett. B \textbf{373}, 282 (1996).
\bibitem{sing} Singh, J. P. and Baghel, P. S.: Elect. J. Theor. Phys. \textbf{6}, 85 (2009).
\bibitem{verm} Verma, M. K., Zeyauddin, M. and Ram, S.: Rom. J. Phys. \textbf{56}, 616 (2011).
\bibitem{joh} Johri, V. B., Desikan, K.: Gen. Rel. Grav. \textbf{26}, 1217 (1994).
\bibitem{shr} Ram, S., Singh, C. P.: Int. J. Theor. Phys. \textbf{254}, 143 (1997).
\bibitem{sing2} Singh, G. P., Beesham, A.: Aust. J. Phys. \textbf{52}, 1039 (1999).
\bibitem{red} Reddy, D. R. K., Naidu, R. L., Rao, V. U. M.: Int. J. Theor. Phys. \textbf{46}, 1443 (2007).
\bibitem{adh} Adhav, K. S., Ugale, M. R., Kale, C. B., Bhende, M. P.: Int. J. Theor. Phys. \textbf{48}, 178 (2009).
\bibitem{sin4} Singh, J. K., Sharma, N. K.: Astrophys. Space Sci. \textbf{327}, 293 (2010).
\bibitem{sin5} Singh, J. K.: Mod. Phys. Lett. A \textbf{25}, 2363 (2010).
\bibitem{sin6} Singh, J. K.: Int. J. Mod. Phys. A \textbf{25}, 3817 (2010).
\bibitem{rao}Rao, V. U. M. and Sudha V. Jaya: Astrophys. Space Sci. \textbf{357}, 76 (2015).
\bibitem{sha1}M. Sharif and S. Waheed: Eur. Phys. J. C \textbf{72},  1876 (2012).
\bibitem{sha2}M. Sharif and S. Waheed: arxiv:1211.3795v1.
\bibitem{sen}Sen, S. and Seshadri, T.R.: Int. J. Mod. Phys. D \textbf{12}, 445 (2003).
\bibitem{col} Collins, C. B., Hawking, S. W.: Astrophys. J. \textbf{180}, 317 (1973).
\bibitem{ber1} Berman, M. S.: Nuovo  Cimento \textbf{B 74}, 182 (1983).
\bibitem{hut} Huterer, D., Turner, M. S.: Phys. Rev. D \textbf{64}, 123527 (2001).
\bibitem{chev} Chevallier, M., Polarski, D.: Int. J. Mod. Phys. D \textbf{10}, 213 (2001).
\bibitem{lind} Linder, E. V.: Gen. Rel. Grav. \textbf{40}, 329 (2008).

\end{thebibliography}
\end{document}